\documentclass[aps,prl,amsmath,amssymb,twocolumn,superscriptaddress]{revtex4-1}

\usepackage{graphicx}
\usepackage{dcolumn}
\usepackage{bm}
\usepackage[dvipsnames]{xcolor}
\usepackage{soul}
\newcommand{\CR}[1]{{\color{black} #1}}

\begin{document}

\title{Nonlinear dynamics of laser-generated ion-plasma gratings: a unified description}

\author{H. Peng}\email{penghao1028311@gmail.com}
\affiliation{Science and Technology on Plasma Physics Laboratory, China Academy of Engineering Physics, Mianyang 621900, China}
\affiliation{Department of Optical Science and Engineering, Fudan University, Shanghai 200433, China}
\affiliation{LULI, Sorbonne Universit\'e, CNRS, \'Ecole Polytechnique, CEA, 75252 Paris, France}
\author{C. Riconda}
\affiliation{LULI, Sorbonne Universit\'e, CNRS, \'Ecole Polytechnique, CEA, 75252 Paris, France}
\author{M. Grech}
\affiliation{LULI, Sorbonne Universit\'e, CNRS, \'Ecole Polytechnique, CEA, 75252 Paris, France}
\author{J.-Q. Su}
\affiliation{Science and Technology on Plasma Physics Laboratory, China Academy of Engineering Physics, Mianyang 621900, China}
\author{S. Weber}
\affiliation{ELI-Beamlines, Institute of Physics of the Czech Academy of Science, 18221 Prague, Czech Republic}
\affiliation{School of Science, Xi'an Jiaotong University, Xi'an 710049, China}

\date{\today}

\begin{abstract}
Laser-generated plasma gratings are dynamic optical elements for the manipulation of coherent light at high intensities, beyond the damage threshold 
of solid-stated based materials. 
Their formation, evolution and final collapse require a detailed understanding.  In this paper, we present a model to explain the
nonlinear dynamics of high amplitude plasma gratings in the spatially periodic ponderomotive potential generated by two identical counter-propagating lasers. 
Both, fluid and kinetic aspects of the grating dynamics are analyzed.
It is shown that the adiabatic electron 
compression plays a crucial role as the electron pressure may  reflect the ions from the grating and induce the grating to break in an X-type manner. 
A single parameter is found to determine the behaviour of the grating and distinguish
three fundamentally different regimes for the ion dynamics: completely reflecting, partially reflecting/partially passing, and  crossing. 
Criteria for saturation and life-time of the grating as well as the effect of finite ion temperature are presented. 
\end{abstract}

\maketitle

{\it Introduction.} 
Plasma optical elements are gaining increasing importance for the manipulation of coherent light from high-power lasers. This is due to the much higher fluences 
plasmas can support compared to solid-state optical devices. However, plasmas are dynamic entities with a finite lifetime. 
It is therefore important to undestand in detail the generation, transient phase and saturation mechanisms of plasma-based optical elements. 

Generating quasi-neutral gratings by intersecting laser pulses in under-dense plasmas or at the plasma surface has been proposed leading to many interesting applications 
\cite{Andreev2002,Sheng2003,Wu2005,Andreev2006,Thaury2007,Michel2014,Monchoce2014,Turnbull2016,Turnbull2017,Leblanc2017,Lehmann2016,Lehmann2017,Lehmann2018,Kirkwood2018a,Kirkwood2018b,Peng2019}, 
e.g. photonic crystals \cite{Lehmann2016}, polarizers \& waveplates \cite{Lehmann2018}, holograms \cite{Leblanc2017}, surface plasma waves excitation \cite{Monchoce2014}, etc.
These gratings are particularly interesting to manipulate intense lasers up to picosecond duration.
Multi-dimensional PIC (particle-in-cell) simulations predict the existence of gratings and they have been indirectly observed in experiments of strong-coupling stimulated Brillouin scattering (sc-SBS) 
 amplification\cite{Lancia2010,Lancia2016,JRM,Peng2018}.
 However, up to now, an in-depth understanding of the growth and saturation of plasma gratings, supported by analytical model, is still lacking. 
 While early studies identifed the important role of ion nonlinearities and X-type wavebreaking in the saturation of the
ion fuctuations, simplified model equations were used and the existence of a driver was not considered 
\cite{Andreev2006,Forslund1975a,Forslund1975b, Forslund1979}. More recent studies have emphasized the importance of the driver
on the electrons, while imposing quasi neutrality for the plasma fuctuations and ballistic ions, and including electron temperature effects 
 in an isothermal way \cite{Sheng2003,Lehmann2016,Lehmann2017,Lehmann2018}. 
The isothermal electron response is appropriate 
 in the limit where the transient gratings are ion-acoustic waves that can be driven to large amplitude either resonantly \cite{Chapman2013, Chapman2014, Michel2014,Turnbull2016,Turnbull2017} or 
 nonresonantly \cite{Friedland2017,Friedland2018}.

\CR{
However, these approaches are not appropriate when considering quasi-neutral gratings exhibiting large density fluctations, 
potentially larger than the critical density $n_{c} = m_e \omega^2/(4\pi e^2)$ (for the driving laser frequency $\omega$),
accessible at moderately high laser intensities over short (100's of fs to 10's of ps) time scales.
In this paper, we develop a fully nonlinear model for such quasi-neutral (ion) gratings.
Our analysis shows that, using an adiabatic model for the electron response in ion modes,
it is possible to obtain a unified description of ion gratings for arbitrary electron temperatures and grating amplitudes. 
This description identifies a single parameter $\mu$, that measures the ratio of the (initial) electron temperature to the ponderomotive potential, 
as fully characterizing the ion grating.
This model allows to deduce clear criteria for the saturation and breaking of the grating as well as to predict the peak density value and size of the gratings.
}\\ 

{\it The model.} 
The nonlinear two-fluid equations including the ponderomotive potential and neglecting the electron inertia are:
\begin{subequations}
\begin{gather}
0=e\frac{\partial \phi}{\partial x}-\frac{1}{n_e}\frac{\partial p}{\partial x}-e\frac{\partial \phi_p}{\partial x},\label{eq:electronMomentum}\\
\frac{\partial^2 \phi}{\partial x^2}=4\pi e(n_e-n_i),\\
\frac{\partial n_i}{\partial t}+\frac{\partial(n_iv_i)}{\partial x}=0,\\
\frac{\partial v_i}{\partial t}+v_i\frac{\partial v_i}{\partial x}=-\frac{e}{m_i}\frac{\partial \phi}{\partial x}.
\end{gather}
\label{eq:fullSet}
\end{subequations}

\noindent The ponderomotive potential $\phi_p=\frac{1}{2}\frac{m_ec^2}{e}a_0^2\cos(2kx)$ is generated by two identical counter-propagating lasers. 
Here $k$ is the laser wave vector in the plasma $k=\tfrac{\omega_0}{c}\sqrt{1-n_0/n_c}$ with $n_0$ the unperturbed plasma density,
and $a_0 = eE/(m_e c\omega_0)$ is the normalized laser field amplitude. 
Adiabatic heating implies $p n_e^{-\gamma}=cst$ with $p=n_eT_e$ the electron pressure; here  
$\gamma=3$ is the adiabatic index for one degree of freedom. 
We then define the thermal potential as $\phi_{th}=\frac{3}{2}\frac{T_{e0}}{e}(\frac{n_e}{n_0})^2$, where $T_{e0}$ is the initial electron temperature, and 
Eq.~\eqref{eq:electronMomentum} reduces to the equation on the electrostatic potential $\phi=\phi_{p}+\phi_{th}$. 
Upon normalizing with $x_{unit}=\frac{1}{2k}$, $\quad t_{unit}=\sqrt{\frac{1}{2}\frac{m_i}{m_e}}(kv_0)^{-1}$, 
$v_{unit}=\sqrt{\frac{1}{2}\frac{m_e}{m_i}}v_0$, $\quad n_{unit}=n_0$, Eqs (\ref{eq:fullSet}) reduce to:
\begin{subequations}
\begin{gather}
\mu \frac{\partial^2 n_e^2}{\partial x^2} -\cos x=\nu(n_e-n_i),\label{eq:simplifiedPoisson}\\
\frac{\partial n_i}{\partial t}+\frac{\partial(n_iv_i)}{\partial x}=0,\\
\frac{\partial v_i}{\partial t}+v_i\frac{\partial v_i}{\partial x}=\sin x-\mu\frac{\partial n_e^2}{\partial x}.
\label{eq:normalizedIonVelocity}
\end{gather}
\label{eq:simplifiedFullSet}
\end{subequations}

\CR{\noindent There exists two governing parameters in this model. The first is ${\nu=\omega_{p0}^2/(2 k^2 v_0^2)}$ (with ${v_0 = a_0\,c}$ the electron quiver velocity in the nonrelativistic laser field and $\omega_{p0}$ the initial electron plasma frequency) that defines the transition from electron ($\nu < 1$) to ion ($\nu>1$) gratings. 
Throughout this work, we focus on quasi-neutral ion gratings for which $\nu \gg 1$ [note also that the regime $\nu<1$ (superradiant regime \cite{Shvets1998}) would require considering the electron inertia].
The second parameter is ${\mu=3T_{e0}/(m_ev_0^2)}$, and it is the most important parameter for this study as it completely describes the dynamics of ion gratings.
Last, we should stress that ensuring large amplitude ion gratings requires to operate in the so-called strong coupling regime of stimulated Brillouin scattering.
This requires $(v_0/v_{th})^2>4k_0c_s\omega_{p}/\omega_{p0}^2$ \cite{Forslund1975a},
correspondingly $\mu<\mu_{tr}=[2a_0^2\frac{m_e}{m_i}(\frac{n_c}{n_0})^2(1-\frac{n_0}{n_c})]^{-1/3}$. 
In all situations of interest we are in this regime.}\\

{\it Solution and comparison with kinetic simulations.} 
The set of  Eqs \ref{eq:simplifiedFullSet} was solved for a large range of parameters $\mu$ and $\nu > 1$ and systematically compared to PIC simulations.  Notice that the dominant 
parameter is  $\mu$ and the system is only weakly dependent on $\nu$, as we verified numerically. The main result of this comparison is that the fluid model allows to predict with very good accuracy the initial formation, peak value and size of the grating. However  kinetic simulations are mandatory to describe the long time evolution and allow to identify different regimes and the relevant timescales.  
In the following the fluid model is compared to kinetic simulations for a representative case: $\mu=1.5$ and $\nu=59.5$.
In the PIC simulations the laser wavelength ($\lambda_0$) and laser intensity are such that $I\lambda_0^2 = 5 \times10^{15}\, {\rm W/cm^2 \mu m^2}$ ($a_0 \simeq 0.06$) and  
$T_{e0} = 920\, {\rm eV}$,  corresponding to $\mu = 1.5$. Two identical laser pulses, constant in time but  with a  slowly linearly growing front of $10 \pi\omega_{p0}^{-1}$  
cross inside the plasma. The unperturbed plasma density is $n_0 = 0.3n_c$, which corresponds to $\nu=59.5$. 
The ion temperature is set to $T_{i0} = 1\,{\rm eV}$ and since for these parameters  the strong coupling threshold is $\mu_{tr} \simeq 32$ we are well into
 the sc-SBS regime, so that in the initial  stage the thermal potential is much smaller than the ponderomotive potential and can be neglected. The 
 SMILEI \cite{Derouillat2018}  code is used for the 1D3V PIC simulation. The cell size is $\lambda_0/256$, and 50 particles per cell are used with a mass ratio of 
  $m_i/m_e = 1836$ and $Z = 1$. The plasma profile is a 
  $6\lambda_0$  plateau with $2\lambda_0$ vacuum at each side. The plasma length is shorter than the sc-SBS growth length $c\gamma_{sc}^{-1}\simeq 24.4~{\rm \mu m}$.
 Therefore there is little energy exchange between the lasers \cite{Chiaramello2016,Amiranoff2018}. 
  
\begin{figure}
\centering
\includegraphics[width=0.3\textwidth]{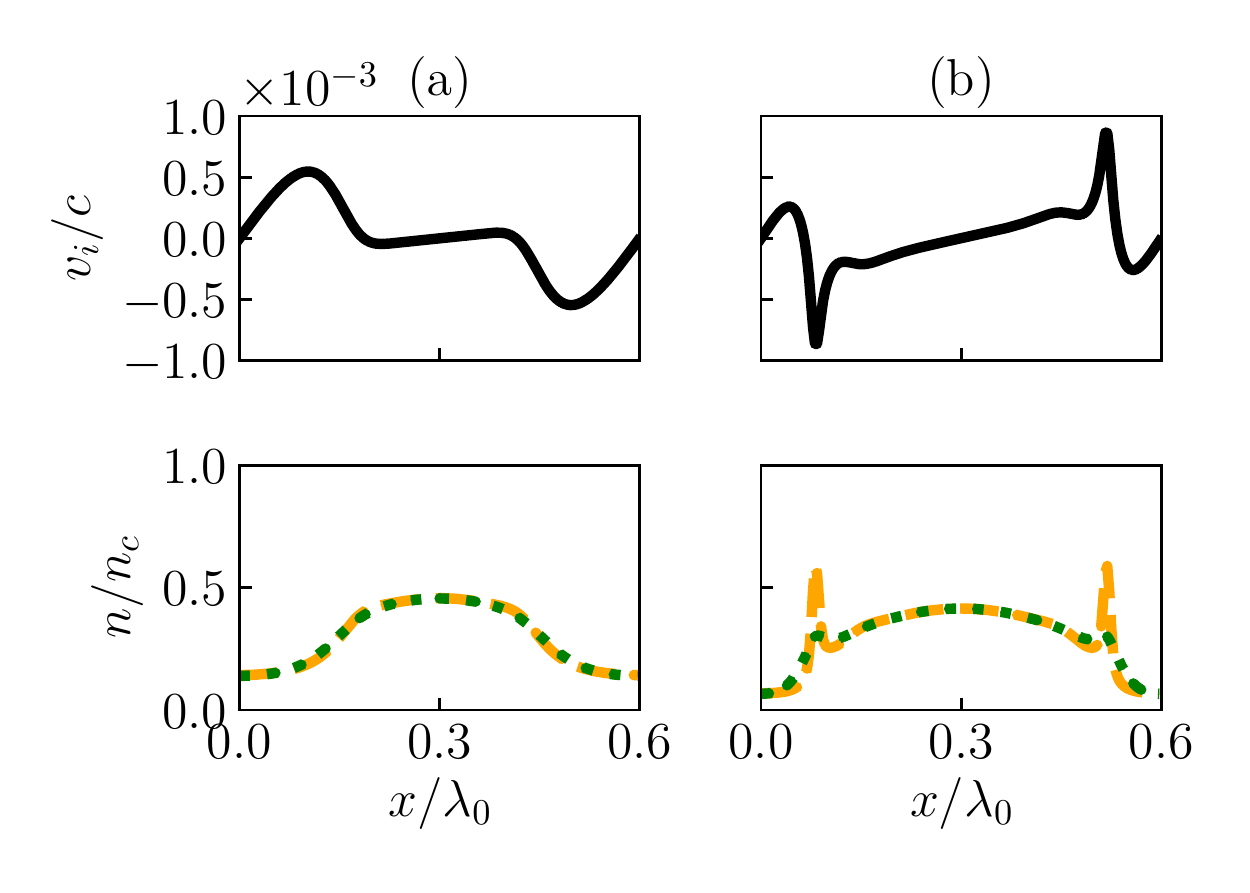}\\
\includegraphics[width=0.5\textwidth]{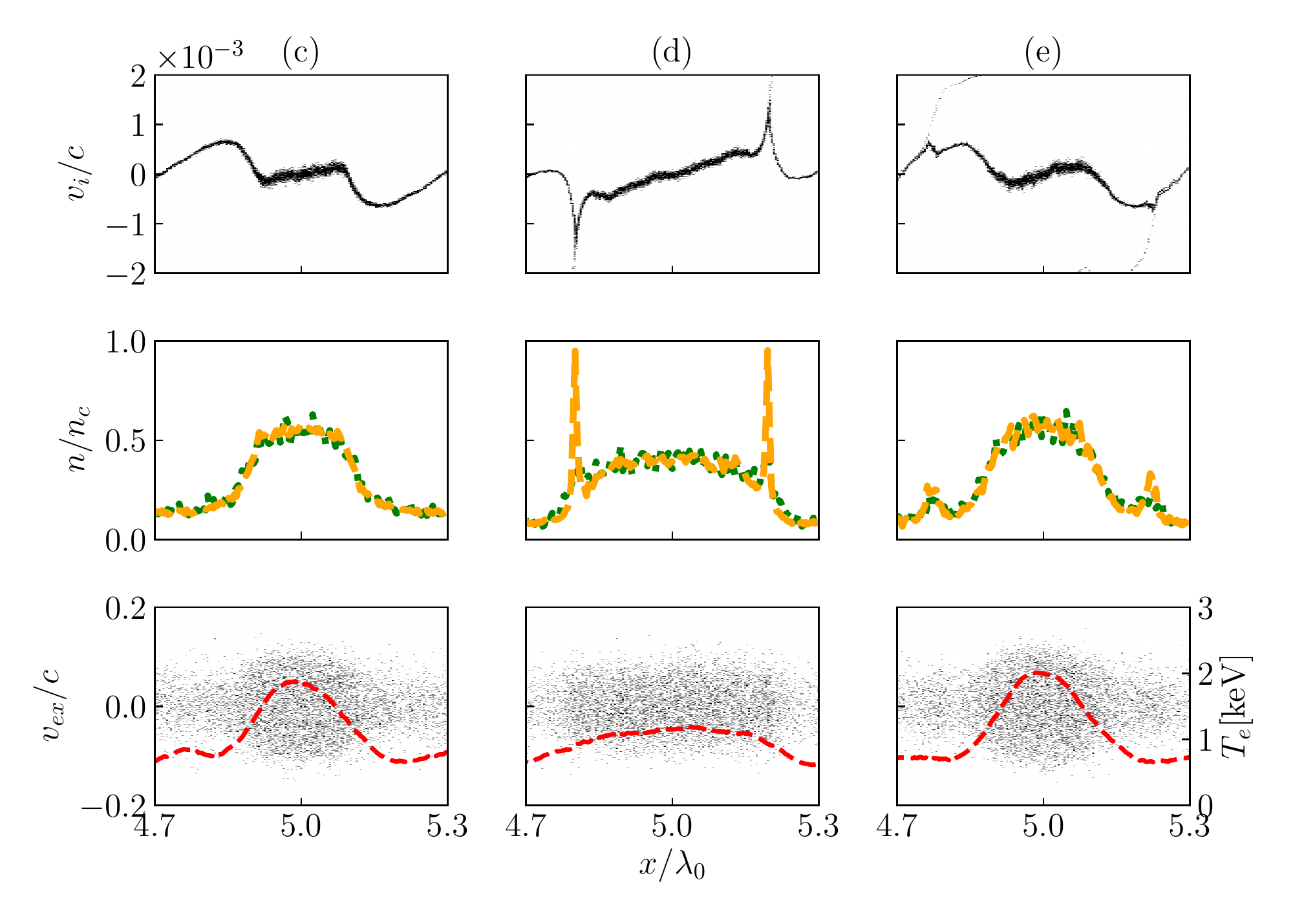}
\caption{Comparison of the fluid model with kinetic simulations. a) and b) show the ion velocity and density profiles for $t = 500\, {\rm fs}$ and  $t = 800\,{\rm fs}$, respectively, 
for $\mu = 1.5$ and $\nu = 59.5$. Displayed are the ion density (green broken line) as well as the electron density (yellow broken line). c), d) and e) present kinetic results for the times $t = 500\,{\rm fs}$, 
$t = 800\,{\rm fs}$ and $t = 1300\,{\rm fs}$, respectively. Also shown (red line) the electron temperature $T_e$ superimposed on the longitudinal phase space.}
\label{fig:PIC}
\end{figure}

The simulation comparison is shown in Fig.~\ref{fig:PIC}, where, as the ponderomotive potential is static and periodic along the x-axis, only one period is shown. Let us consider first the fluid case. 
The ion velocity grows and steepens under the effect of the ponderomotive potential $\phi_p$, leading to the accumulation of ions and electrons towards the potential trough 
[see Fig.~\ref{fig:PIC} (a)]. 
The thermal potential $\phi_{th}=\frac{3}{2}\frac{T_{e0}}{e}(\frac{n_e}{n_0})^2$ (not shown here) also grows quickly as the electron density increases in the grating. 
The combined potential, $\phi = \phi_p + \phi_{th}$ then stops the ions moving toward the center of the grating and a velocity plateau forms for the ions. According to the adiabatic law, the electron temperature  rises to $\sim 2.0 \,{\rm keV}$ at this stage. The ions keep accumulating at the two edges of the grating, generating 
two localized spikes in the ion density [Fig.~\ref{fig:PIC} (b))] stopped by the potential barrier,   $m_i v_i^2 < e\Delta\phi$. 
In the spikes the plasma is non neutral and  the electron density at the center of the grating has reached its maximum value.
The same two phases appear  in the kinetic approach,  Fig.~\ref{fig:PIC} (c) and (d). 
The electron phase space and their 
local temperature are plotted in the bottom row (red dashed line). Indeed the electron temperature in the grating Fig.~\ref{fig:PIC} (c) increases from the initial value, $920\, {\rm eV}$,  to about $2\,{\rm keV}$. \CR{The fluid model, for time that correspond to its range of validity,  reproduces very well the grating dynamics. 
For longer time scales the kinetic approach allows to identify the saturation mechanisms and the subsequent dynamics of the gratings. }
As seen in  Fig.~\ref{fig:PIC} (e) the fastest ions are completely reflected, leading to a X-like ion phase space that corresponds to the X-type wavebreaking widely 
observed in previous works \cite{Forslund1975a, Forslund1975b, Forslund1979, Weber2005a, Andreev2006}. Subsequently the reflected ions induce 
the grating to expand and the plasma density in the grating starts to decrease. The electron temperature now decreases  as the grating stretches, until it is compressed again by the pondemorotive potential.
The net effect of this whole process, compression and stretching of the plasma, leads to the ejection of a small amount of ions in opposite directions, 
that have little effect on the plasma grating maximum as will be discussed later.
Notice that as long as $ZT_e/T_i  > 1$ the following analysis holds, i.e. the model fluid equations can be used in order to 
predict  the peak value and size of the generated gratings. Including a larger ion temperature has simply the effect of smoothing the local non-neutral ion density peaks.\\

\begin{figure}
\centering
\includegraphics[width=0.5\textwidth]{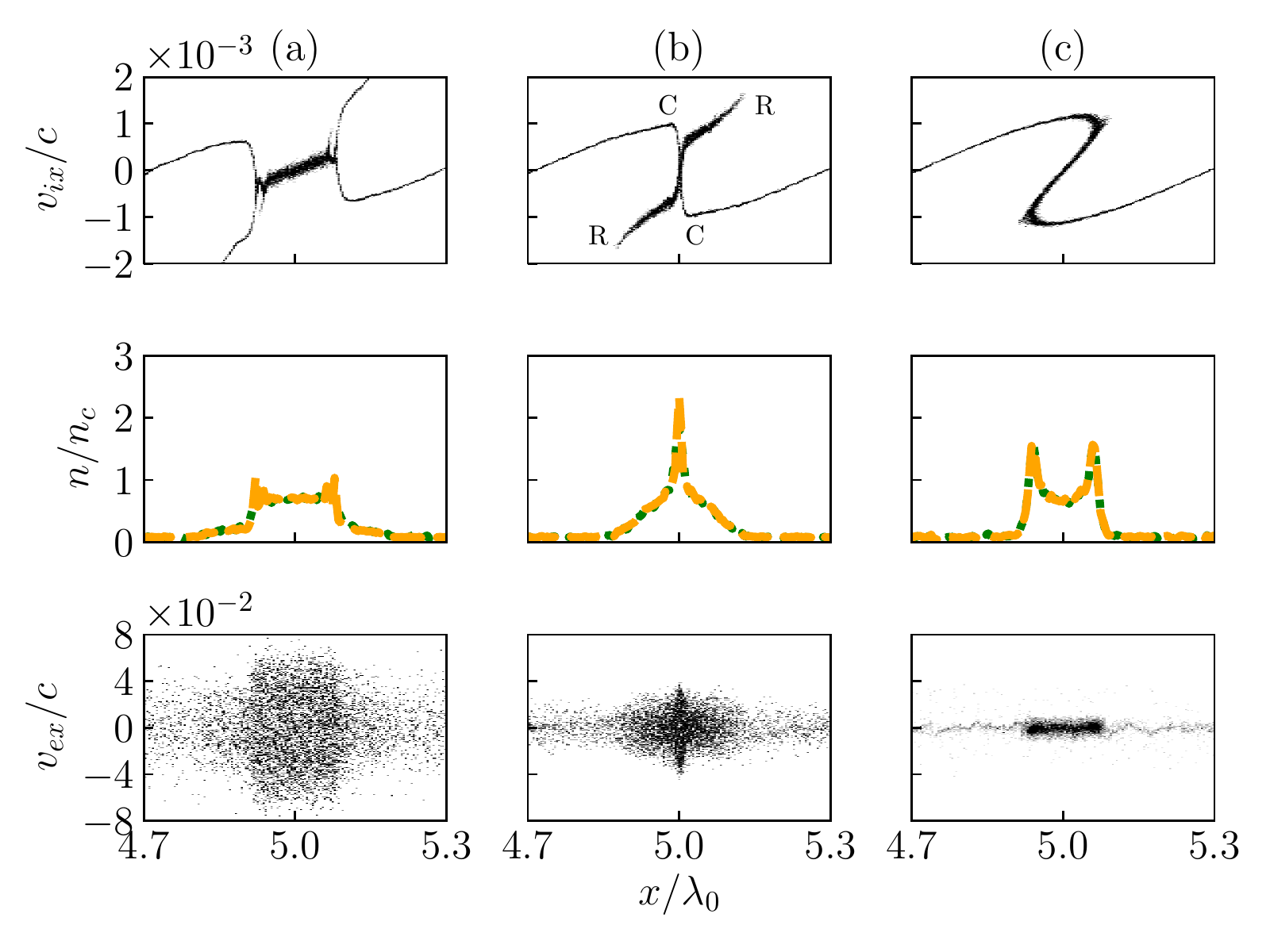}
\caption{ \CR{Three different regimes of the plasma grating: R-regime for $T_{e0} = 153.3 \, \rm{eV} , \mu=0.25 $  (a), T-regime for $T_{e0} = 12.3\,\rm{eV}, \mu=0.02 $  (b), C-regime  for $T_{e0} = 0$ (c) at $t=800\, \rm{fs}$. 
Other parameters are the same as in Fig.(\ref{fig:PIC}). The labels C and R denote crossing and reflecting particles in the phase space.}}
\label{fig:3Regimes}
\end{figure}

{\it Regimes of ion dynamics.} 
The ion kinetic response governs the grating lifetime and the subsequent peaks in the grating. 
Depending on  the parameter $\mu$ one can identify three different regimes for the plasma gratings as function of the ion energy with respect to the ponderomotive potential 
resulting from the electron pressure.
 An understanding of these regimes and an approximate value for the transition can be obtained as follows.  From the model equations one can estimate the maximum kinetic energy acquired by the ions 
  to be of the order of $P_m \sim m_i v_{unit}^2 \sim \frac{m_e v_0^2}{2}$. 
If this kinetic energy is less than the total potential barrier encountered by the ions, they will be reflected and the steepening will stop. This condition corresponds to
 $P_m < e\Delta\phi$, where 
$e\Delta\phi = e\phi_{max}-e\phi(P_m)$, with $\phi_{max}$ the maximum total potential and $\phi(P_m)$  the value of the potential at the position where the ions have
 their maximum energy.  The contribution of the thermal potential to the barrier can be approximated by its maximum value $e (\phi_{th} -\phi_{th}(P_m)) \sim  e \phi_{th}$, \CR{due to the adiabatic heating. 
 Only if it is larger than the absolute value of the ponderomotive potential ($m_e v_0^2/2$) the barrier will be positive and reflection will occur, stopping the density growth.} In dimensionless 
 units this corresponds to the condition $1< \mu (n_e^2/n_0^2)$. If $\mu \gg 1 $  a relatively small compression will be enough to induce ion reflection and result in X-type 
 wavebreaking. By contrast, if $\mu \ll 1$, the cold limit holds where all the available particles are compressed at the center of the ponderomotive potential and eventually cross 
 each other. The transition is thus expected to be at  $\mu \sim 1$.  A more precise limiting value is obtained by kinetic simulations at $\mu = 0.25$.  \CR{We can identify three regimes. For $\mu > 0.25$ there is always 
 complete reflection (R-regime) of the ions: as illustrated in Fig.~\ref{fig:3Regimes} (a) the fastest ions are reflected by the potential barrier, and the density and temperature reach a plateau with a finite 
 lifetime.  
 The opposite extreme situation is found  for $\mu<0.001$, illustrated in Fig.~\ref{fig:3Regimes} (c). In this case the potential is never large enough to
  reflect the bulk of the ions that just oscillate in the potential well of the ponderomotive force, crossing each other at the bottom. Only very few slow particles get reflected and do not contribute to the 
  subsequent dynamics. As a result the density reaches a very large value, but after the particles crossing (C-regime), the density drops subsequently. 
  For $0.25>\mu>0.001$, one encounters a transition regime (T-regime) where the fastest ions are first reflected, but at later times, when the potential decreases, ions 
  are still fast enough to cross the new potential barrier. This intermediate situation is shown in Fig.~\ref{fig:3Regimes} (b). One can observe reflected particles (labelled R) and crossing particles (labelled C, 
  at the position $x/\lambda_0=5$) that still have enough energy to overcome the potential barrier and on a longer time scale flatten the peak.}\\

\begin{figure}
\centering
\includegraphics[width=0.5\textwidth]{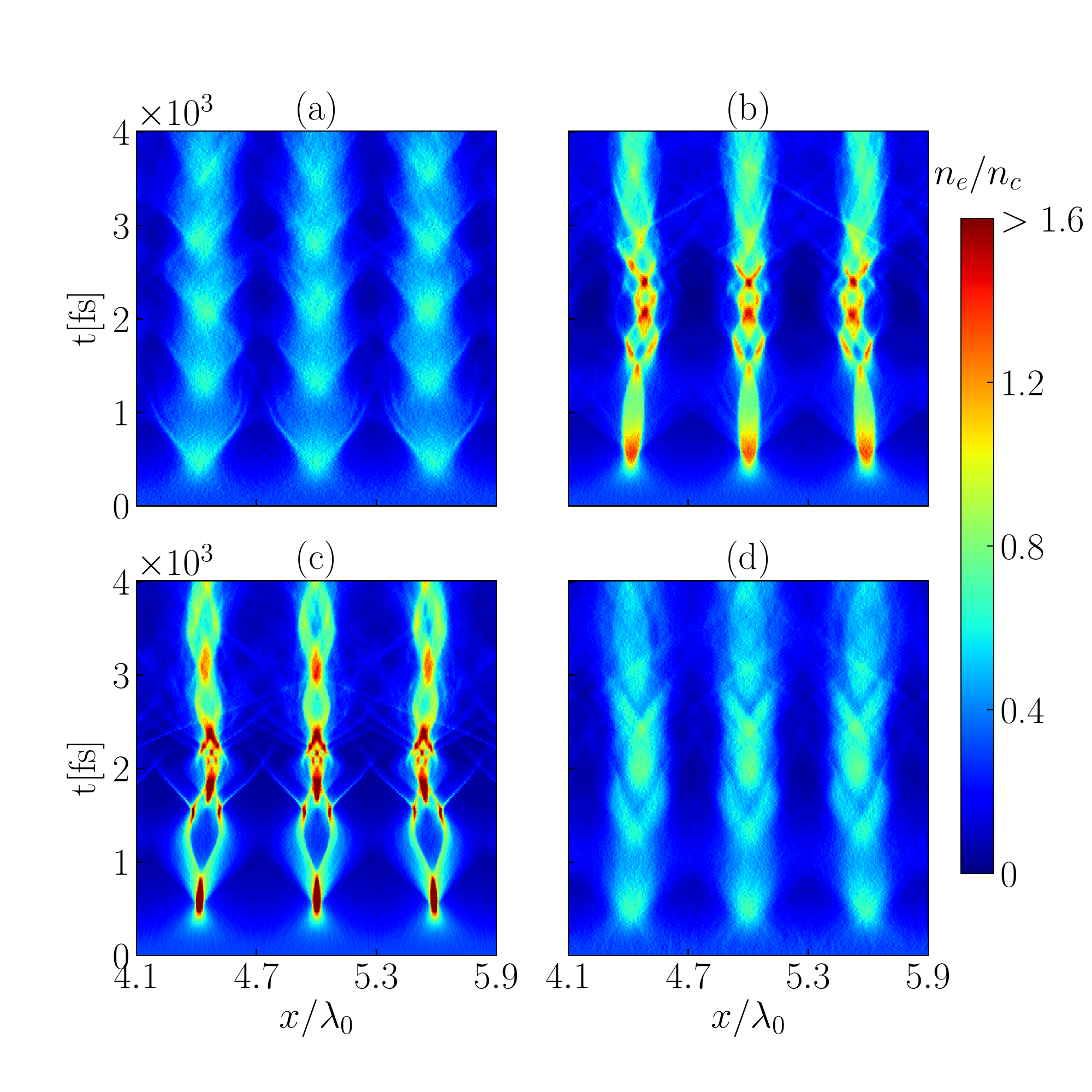}
\caption{Temporal evolution of the electron density at the center of the simulation box (over 3 wavelengths) for $\mu = 1.0$ (a), $\mu = 0.1$ (b) and $\mu = 0.02$ (c). (d) 
shows the effect of the ion temperature with $T_{e0} = T_{i0} = 200\, {\rm eV}$, $\mu = 0.3$.}
\label{fig:time}
\end{figure}

{\it Growth to peak value.} 
The typical time of grating formation as deduced from the fluid equations and confirmed by PIC simulations scales with 
$t_{unit} \propto 1/a_0$ and is given by  $\tau_{form} = 1.5\,t_{unit} = 465 \, {\rm fs}$ in our simulations. 
This value depends weakly on the plasma grating regime, slightly increasing as $\mu$ decreases as shown in Fig.~\ref{fig:time} (a)-(c).  The subsequent evolution instead  depends strongly on the value of $\mu$. 
In the R-regime the  grating periodicity is regular:  the lifetime and the re-generation time is of the same order as the formation time. 
As $\mu$ decreases and the system enters the T-regime the value of the first peak increases but, as seen in Fig.~\ref{fig:time} (b)-(c) subsequent peaks form later in time and have lower density value, while the electrons undergo some heating. 
In the C-regime, the time for the re-generation of the peak is simply due to the bouncing motion of the ions after crossing at the bottom of the potential well. 
By approximating the well by a parabola,  this is simply given by $1/\omega_{B}= 2 \pi t_{unit} = 1953 \, {\rm fs}$. 
A more precise value can be obtained by PIC simulations for the case $\mu= 0$ as $1250 \, {\rm fs}$. 
In order to increase the lifetime of gratings finite ion temperature effects can be considered. 
This will influence the ion reflection and crossing. 
 Nevertheless, even with finite ion temperatures, the three regimes mentioned above still exist. 
The result with finite ion temperature (in the R-regime) is to diminish the central density peak but increase the lifetime, as shown for example in Fig.~\ref{fig:time} (d).
In this figure the electron temperature is taken equal to $T_{e0} = 200\,eV$ i.e. $\mu=0.3$. 
Finite ion temperature plays the role of a larger $\mu$-value: as we can see the peak density is analogous to the case (a) ($\mu=1$  but $T_{i0}=0$), but the lifetime of the grating is increased. 
An appropriate choice of parameters can even lead to a quasi steady state of the density peak. \\

\begin{figure}
\centering
\includegraphics[width=0.5\textwidth]{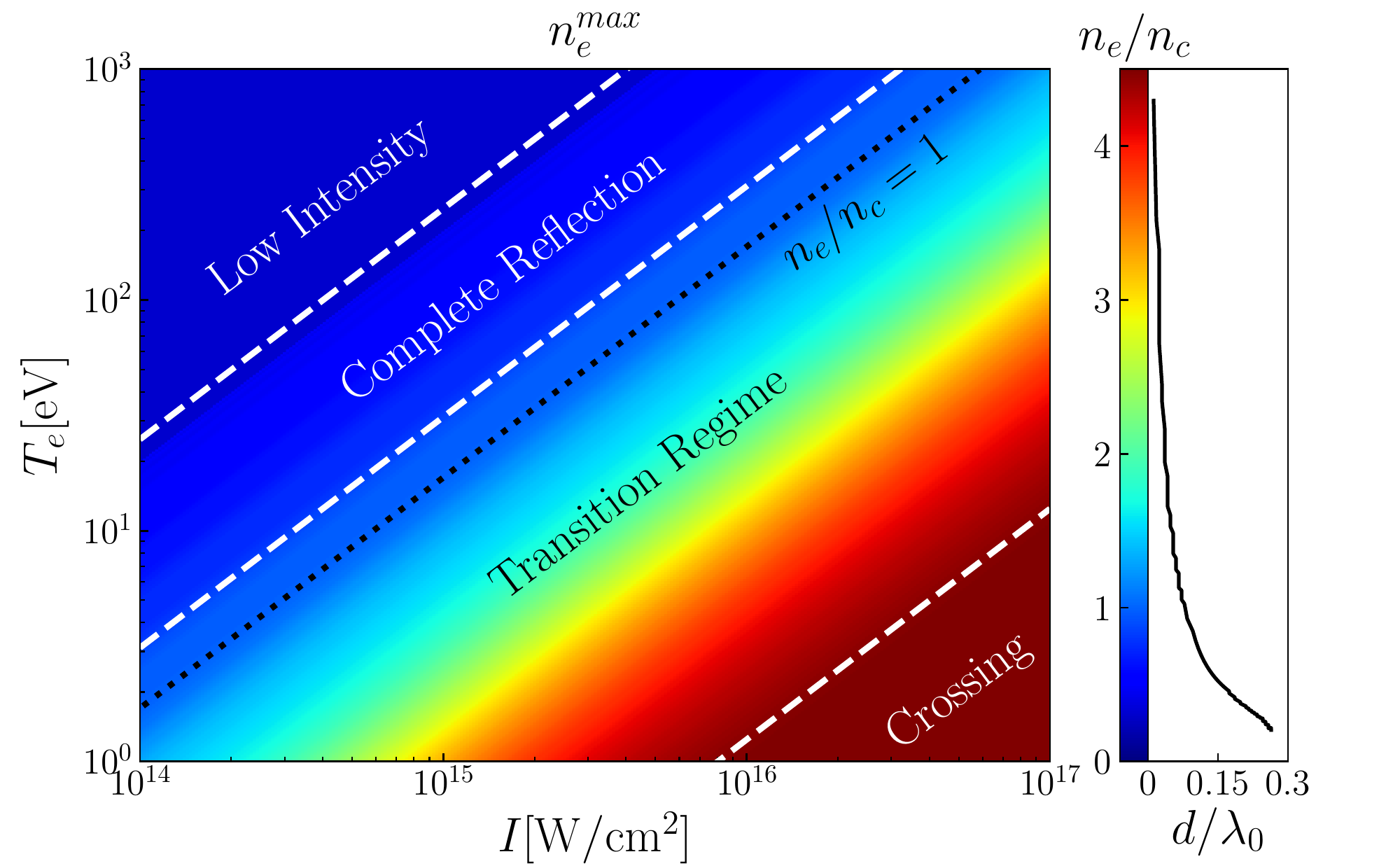}
\caption{Plateau density as a function of the electron temperature and the laser intensity; note that $\mu \propto T_{e0}/I$. The three broken lines present the 
low intensity case ($\mu = 2$), the complete reflection case ($\mu = 0.25$), and the crossing case ($\mu = 0.001$). The right panel shows the minimum grating width $d$ as a function of the plasma 
density of the plateau. The present figure was obtained for a specific density value of $0.3\, n_c$, however the results are generic.}
\label{fig:TeVsI}
\end{figure}

{\it Discussion.} 
\CR{The solution of the fluid equations and the existence of the three regimes are summarized in Fig.~\ref{fig:TeVsI} where we plot the grating peak density for a given set of $I$ and $T_{e0}$ and the minimum grating width $d$ as a function 
of the density of the plateau. Equipotential lines in the figures correspond to values of $I$ and $T_{e0}$ leading to the same $\mu$ and allow to identify the regions of transition among different kinetic regimes.  In general the plateau size is a fraction of the laser wavelength and goes  to zero as $\mu$ decreases and the peak density increases. It is now straighforward to obtain the grating peak density for a given set of $I$ and $T_{e0}$. For example if we consider $a_0= 0.02$  and $T_{e0}=10 eV$ we find a peak density of $n_e/n_c = 0.98$, to be compared with $n_e/n_c = 1.1$ reported in the Ref. \cite{Lehmann2016} as result of a full PIC 2D simulation. }
If the initial density is  lower ($n_e < 0.25\, n_c$) gratings can still be formed 
\cite{Weber2013, Riconda2013,Riconda2015,Chiaramello2016,Chiaramello2016b,Amiranoff2018,Peng2016,Zhang2017}, 
but Raman backscattering leads to the generation of hot electrons which will heat the grating 
and increases the parameter $\mu$. \CR{At very low temperature the role of collisions in principle needs to be taken into account. We verified by simulations that predictions from the fluid model  still hold for temperatures as low as 10 eV,  nevertheless for longer time scales the ratio between $t_{unit}$ and the typical collision time has to be considered and collisions might be included in the kinetic model to properly describe the lifetime and evolution of the grating.} 
Also note that the validity of the above discussion and Fig.~\ref{fig:TeVsI} resides on the assumption that the driver overlap time is at least of the order of the characteristic 
grating formation time $\tau_{form} \propto 1/a_0$. For shorter times the model is still valid but the peak density will be smaller and can be calculated from the fluid equations.
A unified model was presented of the nonlinear dynamics and saturation of ion-plasma gratings generated by two driving laser beams in a 
self-consistent way. \CR{This provides the tools to dimension the gratings for the required application \cite{Andreev2002,Sheng2003,Wu2005,Andreev2006,Thaury2007,Michel2014,Monchoce2014,Turnbull2016,Turnbull2017,Leblanc2017,Lehmann2016,Lehmann2017,Lehmann2018,Kirkwood2018a,Kirkwood2018b,Peng2019} as plasma gratings for the manipulation of coherent light can be generated in 
a controlled way and fine-tuned for a specific purpose. This is another important example of the use of laser-modulated plasmas for high-power laser science and the possibility to control light by light.}\\

The authors wish to acknowledge useful discussions with F. Amiranoff and J. Wurtele.
This work has been done within the LABEX Plas@par project, and received financial state aid managed by the Agence Nationale de la Recherche, as part of the 
program "Investissements d'avenir" under the reference ANR-11-IDEX-0004-02.
H.P acknowledges the funding from China Scholarship Council, and is partially founded by the Natural Science Foundation of China and 11875240 and Key lab foundation of CAEP 6142A0403010417. S.W. was supported by the project Advanced research using high intensity laser 
produced photons and particles (ADONIS) (CZ.02.1.01/0.0/0.0/16\_019/0000789)  and by the project High Field Initiative 
(HiFI) (CZ.02.1.01/0.0/0.0/15\_003/0000449), both from European Regional Development Fund. 
The results of the  Project LQ1606 were obtained with the financial support of the Ministry of Education, Youth and Sports as part of targeted support from the National Programme of Sustainability II (SW).

\end{document}